\documentclass[aps,prb,twocolumn,showpacs,floatfix,amsmath,amssymb,
               superscriptaddress]{revtex4-1}
\usepackage{color}
\usepackage{graphicx}
\usepackage{natbib}
\usepackage{epsfig}
\usepackage{setspace}
\usepackage{amsmath,amssymb}
\usepackage{verbatim}
\usepackage{bibentry}

\def\etal{{\em{et al. }}}
\def\CPA{\scriptscriptstyle CPA}

\newcommand{\ep}{\ensuremath{\epsilon}}

\newcommand{\epk}{\ensuremath{\epsilon(\mathbf{k}^{||})}}
\def\nnu{{\nonumber}}

\begin{document}
\title{Spectral changes in layered $f$-electron systems induced by 
Kondo hole substitution in the boundary-layer}
\author{Sudeshna Sen}
\affiliation{Chemistry and Physics of Materials Unit, JNCASR,
Bangalore-560064, Karnataka, India}
\author{J.\ Moreno}
\author{M.\ Jarrell}
\affiliation{Department of Physics \& Astronomy, Louisiana State University,
Baton Rouge, LA 70803, USA}
\affiliation{Center for Computation and Technology, Louisiana State University, Baton Rouge, LA 70803, USA}
\author{N. S. Vidhyadhiraja}\email{raja@jncasr.ac.in}
\affiliation{Theoretical Sciences Unit, JNCASR,
Bangalore-560064, Karnataka, India}
\affiliation{Department of Physics \& Astronomy, Louisiana State University,
Baton Rouge, LA 70803, USA}

\begin{abstract}
\noindent  We investigate the effect of disorder on the dynamical spectrum of 
          layered $f$-electron systems. With random dilution 
          of $f$-sites in a single Kondo insulating layer, we explore the 
          range and extent to which Kondo hole incoherence can penetrate into 
          adjacent layers. We consider three cases of neighboring layers: 
          band insulator, Kondo insulator and simple metal. 
          The disorder-induced spectral weight transfer, used here for 
          quantification of the proximity effect, decays algebraically with 
          distance from the boundary layer. Further, we show that the spectral 
          weight transfer is highly dependent on the frequency 
          range considered as well as the presence of interactions in the 
          clean adjacent layers. The changes in the low frequency
          spectrum are very similar when the adjacent layers are either 
          metallic or Kondo insulating, and hence are independent of 
          interactions. In stark contrast, a distinct picture emerges
          for the spectral weight transfers across large energy scales. 
          The spectral weight transfer over all energy scales is much higher 
          when the adjacent layers are non-interacting as compared to 
          when they are strongly interacting Kondo 
          insulators. Thus, over all scales, interactions screen 
          the disorder effects significantly. We discuss the possibility of a 
          crossover from non-Fermi liquid to Fermi liquid behavior upon 
          increasing the ratio of clean to disordered layers in 
          particle-hole asymmetric systems.
\end{abstract}
\maketitle
\section{INTRODUCTION}
Several surprises have emerged through extensive experimental and 
theoretical investigations of layered correlated systems over the last decade.
The pioneering study of Ohtomo \etal led to the discovery of a two-dimensional
 electron gas at the interface of Mott insulating LaTiO$_3$ and band insulating SrTiO$_3$ in atomically resolved 
heterostructures \cite{Ohtomo2002}. 
A dimensional driven crossover from metal to 
insulator transition ~\cite{MIT_thinfilms_2010} 
and an anomalous effective mass enhancement ~\cite{SrVO3_thinfilms}
was observed by Yoshimatsu \etal in digitally 
controlled SrVO$_3$ thin films. Theoretical predictions in this regard 
had been reported and semiquantitatively explained by Okamoto \etal 
in Ref.~\onlinecite{Okamoto_2008}. An early work in this regard may also 
be found in Ref.~\onlinecite{Pothoff_effective_mass}.
In general, strongly 
correlated interfaces exhibit several other unexpected properties such as 
superconductivity \cite{Reyren2007}, coexistence of ferromagnetism and 
superconductivity \cite{Bert2011}, and electronic phase separation 
\cite{Wang2011}. 
Such investigations have led to a renewed focus on
a number of intriguing aspects of layered systems. Some of these issues are due to atomic
reconstruction, enhanced correlation effects due to reduced coordination 
number and emergent energy scales owing to the presence of disparate orbital, 
charge, and spin degrees of freedom \cite{hetero_rev}.
    
A significant number of theoretical studies 
on the proximity effects of electron-electron interaction in layered 
systems have been carried out. 
Such studies include proximity effects in Hubbard layers 
\cite{Nolting1999, *Nolting2003, *MIT_thinfilms, *surface_MIT_Pothoff,
*largeU_Pothoff1996, Millis2004, *Okamoto2005, *Millis2005,
fragileFL, Scalettar2012, embedding_Ishida, Ishida_CDMFT_2010, 
Ishida_2012, Helmes2008, Ruegg1, *Ruegg2}, 
Falicov-Kimball layers \cite{IDMFT2004} 
or $f$-electron superlattices \cite{Peters1, Peters2}. 
Zenia \etal~\cite{fragileFL} demonstrated that a Mott insulator transforms to a 
"fragile" Fermi liquid if sandwiched between metallic leads. 
Helmes \etal~\cite{Helmes2008} studied  interfaces of strongly correlated 
 metals with Mott insulators. 
The rate of decay of the quasiparticle weight  
into the Mott insulator was quantified and further corroborated in 
the spirit of a Ginzburg-Landau mean field treatment. 
Ishida and Liebsch~\cite{Ishida_CDMFT_2010} investigated  the effect
of an interplanar Coulomb interaction using the
cellular Dynamical Mean Field Theory (DMFT) and observed a 
non-local correlation induced reduction of the proximity effect. 
Cluster extensions of DMFT was employed by Okamoto \etal in 
Ref.~\onlinecite{Okamoto_Maier} to examine the proximity effect 
in superlattices involvng cuprates, predicting a novel enhancement 
in the superconducting transition temperature. 

Recent experimental investigations of heavy fermion ($f$-electron) 
superlattices  indicate a fascinating interplay of heavy fermion 
physics, low dimensionality and interface effects 
\cite{dimensional_confinement2010}. 
In the heavy fermion superlattices $m$CeIn$_3$-$n$LaIn$_3$, 
by reducing the thickness 
of the CeIn$_3$ layers grown on metallic LaIn$_3$, it was 
demonstrated that the dimensionality of the $f$-electrons and the 
magnetic order could be controlled. These dimensionally confined heavy fermion systems
then displayed non-Fermi liquid properties that manifested as a linear 
temperature dependence in resistivity, $\rho_{xx}\sim T$, with a $T^2$ 
behavior recovered on increasing the number of CeIn$_3$ 
layers. This has been followed up with very recent theoretical 
investigations of the Kondo effect and dimensional crossover in 
$f$-electron superlattices \cite{Peters1, Peters2, Peters3}. 
Tada \etal~\cite{Peters2} analyzed the formation of heavy 
electrons in $f$-electron multilayers. They demonstrated the 
existence of two (in-plane and out of plane) coherence temperatures in 
such systems. This implies a crossover in dimensionality of the heavy 
fermions from two to three dimensions as the temperature and the geometry 
of the system change.  In a study of interfaces of Kondo lattice layers and 
normal metals, Peters \etal~\cite{Peters1} showed that such a coupling transformed the 
full gap of the Kondo lattice layers into a vanishing soft gap. They also demonstrated 
the strong influence of the Kondo effect on the 
density of states of the metallic layer. This proximity effect was further shown 
to be strongly dependent on the number 
of non-interacting metallic layers. 

Although the proximity effects of strong
interactions have been addressed quite rigorously, the
 effects of disorder have not been considered.
 Incoherent scattering due
 to impurities is inevitable in heterogeneous interfaces and therefore 
the physical effects stemming from disorder could be significant. In fact, 
in heavy fermion compounds,
when magnetic sites are substituted by non-magnetic impurities 
    a substantial reduction of the coherence 
    temperature $T_{coh}$ occurs \cite{CPA_NRG}. Shimozawa \etal studied the effects of Kondo 
hole disorder on epitaxial thin films of divalent-Yb substituted CeCoIn$_5$~\cite{CeCoIn_thin_films}. 
It has been shown in Ref~\onlinecite{CPA_NFL} that dynamical scattering from Kondo holes 
can even yield a non-Fermi liquid behavior  for bulk systems.  Hence, a theoretical 
    study of Kondo hole substituted $f$-electron interfaces is highly relevant. 
    
    In this work, we have explored the spectral dynamics of a single 
    substitutionally disordered Kondo-insulator layer at the boundary of 
    several clean layers of which three 
    types have been considered: (i) non-interacting metals or 
    (ii) band-insulators
    and (iii) strongly interacting Kondo insulators. We have investigated the 
penetration of disorder induced impurity scattering into the proximal 
layers. An interplay of interactions and Kondo hole 
disorder lead to 
significant differences in the spectral weight transfer at low frequencies 
versus the overall spectrum.  We argue that, for non particle-hole symmetric 
systems, the systematic addition of clean interacting layers to the 
substitutionally disordered 
interface could lead to a gradual crossover from non-Fermi liquid to Fermi 
liquid behavior.  The paper is organized as follows: The model and methods
 are described in the next section. The results and discussion follow in Section 3. 
 Our conclusions are presented in the final section.
    
  \section{Model and method}
  \label{sec:model}
  
  The Hamiltonian of a heavy fermion layered system 
  may be constructed through a slight generalization of the standard Periodic Anderson model (PAM) as:
   \begin{align}
    \label{PAM}
   \mathcal{H}
   =&-\sum_{ij\alpha\sigma}t_{i\alpha j\alpha}^{||}
    \left(c_{i\alpha\sigma}^{\dagger}c_{j\alpha\sigma}^{\phantom\dagger}
    +h.c.\right) \nonumber\\   
    &+\sum_{i\alpha\sigma}V_\alpha^{\phantom\dagger}
     \left(f_{i\alpha\sigma}^{\dagger}c_{i\alpha\sigma}^{\phantom\dagger}
    +h.c.\right)\nonumber\\ 
    &+\sum_{i\alpha\sigma}\left(\epsilon_{c\alpha}^{\phantom\dagger}
    c_{i\alpha\sigma}^{\dagger}c_{i\alpha\sigma}^{\phantom\dagger}
    +\epsilon_{f\alpha}^{\phantom\dagger}
    f_{i\alpha\sigma}^{\dagger}f_{i\alpha\sigma}^{\phantom\dagger}\right)+ \nonumber\\
    &+\sum_\alpha U_\alpha n_{fi\alpha\uparrow}n_{fi\alpha\downarrow}
    -\sum_{i\alpha\sigma}t^{\perp}_{\alpha\alpha+1}
    \left(c_{i\alpha\sigma}^{\dagger}c_{i\alpha+1\sigma}^{\phantom\dagger}
    +h.c.\right), 
  \end{align}
  where, $t_{i\alpha j\alpha}^{||}$ represents the in-plane hopping 
  between the conduction band ($c$-electron) orbitals at sites $i$ and $j$, 
  in the plane $\alpha$; $\epsilon_{c\alpha}$ and $\epsilon_{f\alpha}$ are 
  the onsite energies of the $c$- and  $f$-electrons, respectively; 
  $V_\alpha$ is the hybridization between the heavy $f$-electrons and 
  the $c$-electrons in the plane $\alpha$; $U_\alpha$ is the onsite Coulomb 
  repulsion between two electrons occupying an $f$ orbital in the plane 
  $\alpha$; and $t^{\perp}_{\alpha\alpha+1}$ represents the interplane 
  hopping between the delocalized $c$-orbitals. We explicitly assume here 
  that the $f$-orbitals being local have negligible overlap between the 
  layers.
  
  The layer Green's functions, $G^{cc}_{\alpha\beta}$ for 
  the $c$-orbitals, may be obtained through an equation of motion method,
  applied in real space \cite{IDMFT2004}.  For an $N$-layered system,
  the entire matrix of the c-Green's functions is given by the following expression:
   \begin{align}
\label{locGcc}
    &\mathbf{G}^{cc}(\omega,\mathbf{k}^{||})\\ \nonumber
     &=
     \begin{pmatrix}
      \lambda_1-\epk & -t_{\perp} & 0 & \hdots \\
      -t_{\perp} & \lambda_2-\epk & -t_{\perp} & \hdots \\
      0 & -t_{\perp} & \lambda_3-\epk & \hdots \\
      \vdots & \vdots & \vdots & \vdots & \vdots \\
      0 & 0 & 0 & \lambda_N-\epk \\
    \end{pmatrix}^{-1}, 
   \end{align}
   with $\lambda_\alpha=\omega^{+}
         -\epsilon_{c\alpha}-\Sigma_{c\alpha}$, $\omega^{+}= \omega + i \delta$, and $\mathbf{k^{||}}$ 
   denotes the Bloch vector along the planar direction. 
   Equation \eqref{locGcc} assumes the existence of translational invariance 
   along the in-plane direction. The term $\Sigma_{c\alpha}$ is the 
   $c$-selfenergy arising due to the hybridization with the correlated  
   $f$-orbitals. The above expression is not restricted to the PAM,
    and can be used in more general situations, e.g. if
   the $c$-orbitals were correlated and thus had an intrinsic selfenergy.
   The $\lambda_\alpha$ are determined by the Hamiltonian of the 
   $\alpha^{th}$ layer. 
   Note that the band insulator is constructed 
   using a fictitious localized orbital that plays no role except 
   to hybridize with the conduction orbital and create a band gap. 
   For example, $\lambda_\alpha$ is given by $\omega^+-\frac{V^2_\alpha}{\omega^+}$,
   $\omega^+$ and $\omega^+ - V_\alpha^2/(\omega^+ - \ep_f
   -\Sigma_{f\alpha})$ for
   band-insulators, metals and Kondo insulators respectively, with 
   $\Sigma_{f\alpha}$ being the local $f$-selfenergy in the $\alpha^{\rm th}$
   layer. Within dynamical mean field theory, where the local approximation
   is valid, an impurity solver is needed to obtain the $f$-selfenergy.
   We have employed the local moment approach
   to solve the effective self consistent impurity problem.
   
   The local $c$-Green's functions are given by a $\mathbf{k^{||}}$ 
   summation of
   the $\mathbf{G}^{cc}(\omega,\mathbf{k}^{||})$ matrix. In order to avoid the 
   numerically expensive step of summation over $k_x,k_y$ for each frequency 
   $\omega$, we employ the method used in Ref.~\onlinecite{Helmes2008}. If the 
   inverse c-Green's function matrix  
   (in Eq.~(\ref{locGcc})) is denoted as $M^{\prime}$, such that 
   $M^{\prime}=M-\epsilon(\mathbf{k^{||}})\mathbb{I}$,  
   a similarity transformation may be used to transform the 
   $M$ matrix into diagonal form, such that,
\begin{equation}
\label{eq:finalgc}
  \mathbf{G}^{cc}(\omega)=
   \mathcal{S}
     \begin{pmatrix}
       H[\Gamma_1] & 0 & 0 & \hdots\\
       0 & H[\Gamma_2] & 0 & \hdots\\
       0 & 0 & H[\Gamma_3] & \hdots\\
       \vdots & \hdots & \vdots & \hdots
     \end{pmatrix}
     \mathcal{S}^{-1},
\end{equation}
where the Hilbert transform,
  $H[z]=\int^\infty_{-\infty}\,d\epsilon\rho_0(\epsilon)
(z-\epsilon)^{-1}$ over $\rho_0(\epsilon)$, the non-interacting planar 
density of states, represents the result of the $\mathbf{k^{||}}$ summation and 
$S$ is the transformation matrix diagonalizing $M$. The $N$ 
eigenvalues of $M$ are denoted by $\Gamma_r$ with $r=1\cdots N$. 
The above procedure
is valid as long as the band-dispersion is the same for every layer.
Within layer-DMFT, each layer is treated as a single-impurity embedded
within a non-interacting host. Thus the $c$-Green's function for the
$\alpha^{\rm th}$ layer may be written as 
\begin{align}
   G_{\alpha,\alpha}^{cc}(\omega)&=\sum_{r=1}^N S_{\alpha r}\,
   H[\Gamma_r]\, (S^{-1})_{r\alpha} \label{eq:ssinv}
\end{align}
\begin{align}
   &= \frac{1}{\lambda_\alpha(\omega) - \Delta_\alpha(\omega)}, 
  \label{eq:FSE}
\end{align}     
where $\lambda_\alpha(\omega)=\omega^{+}-
              \epsilon_{c\alpha}-\Sigma_{c\alpha}$ and
  $\Delta_\alpha(\omega)$ is the host hybridization for the $\alpha^{\rm th}$ PAM layer. Note that 
Eq.~(\ref{eq:ssinv}) and (\ref{eq:FSE}) become the definition of $\Delta_\alpha(\omega)$. 

An alternative way to approach this problem is through a Feenberg selfenergy approach \cite{Feenberg, *Economou}. 
Consider any site on the $\alpha^{\rm th}$ layer. 
The $\Delta_\alpha(\omega)$ for this site may be written as a sum of 
self avoiding walks on the entire lattice with the lines representing 
hopping (intralayer given by $-t^{||}$ and interlayer by $t_\perp$), and the 
vertices being the site-excluded Green's functions \cite{Feenberg, *Economou}. 
The definition of the hybridization through 
equations~\eqref{eq:ssinv} and ~\eqref{eq:FSE} is a practical route to 
summing
all these diagrams. The local approximation of DMFT implies that the 
full selfenergy is momentum independent, hence the $\lambda_r$
will get modified if interactions are included on the $r^{\rm th}$ layer.
However, since that is a local change, the diagrams for the $\Delta_\alpha(\omega)$ do not change.This implies that the definition of the host hybridization in the presence
 of interactions remains the same as that for the non-interacting case within
 DMFT, albeit computed with a new set of $\lambda_r$. Including Kondo-hole disorder on the $r^{\rm th}$ layer within the Coherent Potential Approximation (CPA) does not change this definition and is 
 hence tantamount to redefine 
$\lambda_r$ in the following way \cite{CPA_NFL}:
 \begin{equation}
 \frac{1}{\lambda_r(\omega)-\Delta_r(\omega)} = \frac{1-p}{\gamma_r(\omega) - \Delta_r(\omega)} + \frac{p}{\gamma_{0r}(\omega) - \Delta_r(\omega)},
 \label{eq:cpa}
 \end{equation}
where $\gamma_r(\omega)= \omega^+ - \epsilon_{cr}-\Sigma_{cr}$
represents the sites with $f$-electrons and $\gamma_{0r}(\omega)=\omega^+
-\epsilon_{cr}$ represents the Kondo hole sites. Thus, for a given
set of $\Sigma_{cr}(\omega)$, the $\lambda_r(\omega)$ and $\Delta_r(\omega)$
need to be determined self consistently by combining 
equations~\eqref{eq:ssinv} and ~\eqref{eq:FSE} with 
equation~\eqref{eq:cpa}. A practical computational
procedure is the following. Step-1: Guess a set of $\Delta_r(\omega)$
and use them to find $\lambda_r$ from equation~\eqref{eq:cpa}. Step-2:
Use these $\lambda_r$ in equations~\eqref{eq:ssinv} and ~\eqref{eq:FSE} to find
a new set of $\Delta_r(\omega)$. Go back to Step-1 until convergence is achieved.
The obtained $\lambda_r(\omega)$ may be used to define a disorder-averaged
selfenergy for the $r^{\rm th}$ layer as
\begin{equation}
\Sigma^{\CPA}_{cr}(\omega)=\omega^+ - \epsilon_{cr}-\lambda_r. 
\label{eq:sigcpa}
\end{equation}

The self consistent interacting impurity problem is solved here using the local
moment approach (LMA). The LMA is a diagrammatic perturbation theory 
built around the two broken-symmetry, local moment solutions ($\mu=\pm |\mu_0|$) of 
an unrestricted Hartree-Fock approximation. The diagrams that 
incorporate dynamics are the transverse
spin-flip processes. The symmetry broken at the mean-field level is 
restored by imposing adiabatic continuity to the non-interacting limit,
and this step leads to the emergence of a Kondo scale. The details of the
LMA for the single impurity Anderson model and the PAM may be found in references
\cite{Victoria2003, Raja2004, Raja2005, Raja2_2005}. The extension to finite 
disorder may be found in Ref.~\onlinecite{CPA_NFL}. We combine LMA, CPA and the inhomogeneous DMFT \cite{IDMFT2004} to explore the effect of disorder
and interactions in layered-PAM systems.
\section{Results}
\label{sec:results}
We have explored the proximity effects of a Kondo hole-disordered heavy fermion layer 
in thin films, by studying geometries where the substitutionally 
disordered Kondo insulator 
is in proximity to several clean interacting or non-interacting layers. 
Before we delve into results obtained using
the numerical implementation of the local moment approach within inhomogeneous DMFT,
we present a few general results that are exact within this framework of layered systems.

\subsection{Analytical results for a few special cases}
\label{subsec:analytics}
Using the equations detailed in Section~\ref{sec:model}, it is easy to 
obtain closed-form expressions for the Green's functions in certain 
simple cases. We present and discuss a few results for a non-interacting 
layered system,
a bilayer and a trilayer system.

\subsubsection{Non-interacting case}
\label{subsubsec:NIcase}
For $N$-identical layers
the matrix on the right hand side of Eq.~\eqref{locGcc} has the structure of a symmetric tridiagonal 
Toeplitz matrix \cite{toeplitz}. The complete
eigen-spectrum and corresponding eigenvectors for such a matrix are known in 
closed form \cite{toeplitz}, and hence may be used in 
combination with Eq.~\eqref{locGcc} 
to find the exact Green's functions for the $n^{\rm th}$ layer of a system 
of $N$ identical non-interacting metallic layers (see also Ref.~\onlinecite{Potthoff}).
The density of states at a fixed $\omega$ as a function of layer-index becomes:
\begin{equation}
A_n(\omega)=\frac{2}{N+1}\sum_{m=1}^N \sin^2\left(\frac{mn\pi}{N+1}\right)
\rho_0\left(\omega+2t_\perp\cos\frac{m\pi}{N+1}\right)\,.
\label{eq:nonint}
\end{equation} 
With the replacement of $\omega$ by $\omega-V^2/\omega$ in the 
above equation, the result describes $N$ identical non-interacting
band-insulators also.
\begin{figure}[h!]
\centerline{\includegraphics[clip=,scale=0.5]
                        {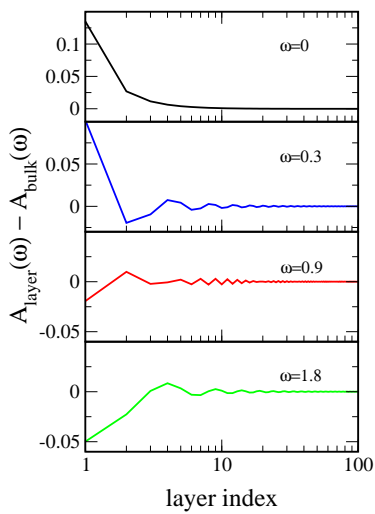}}
\caption{The change in layer density of states (compared to the bulk) as a 
function of layer index, for various frequencies. The result was obtained 
using Equation \eqref{eq:nonint} for a system of 500 layers. The 
bare density of states for each layer $(\rho_0(\epsilon))$ 
is chosen to be of semi-elliptic form,
$\rho_0(\epsilon)=\sqrt{1-\epsilon^2/t_{||}^2}/(2\pi t_{||}).$}
\label{fig:nonint}
\end{figure}
$A_n(\omega)$, shown in Figure~\ref{fig:nonint}, exhibits oscillatory
behavior for all $\omega$ values except  $\omega=0$ where a monotonic decay is observed. 
These well-known surface Friedel oscillations arise
because the Fermi surface of the infinite metallic system has been 
perturbed by the presence
of terminal surfaces \cite{IDMFT2004, Nolting2003}. 

\subsubsection{Bilayer systems: General considerations}
\label{subsubsec:bilayer}
For a bilayer system, the Green's functions are given by
\begin{align}
G_{11}(\omega)&= a\,H[\Gamma_+] +b\,H[\Gamma_-],  \label{eq:bilayer1}
\end{align}
\begin{align}
G_{22}(\omega)& = a^\prime\, H[\Gamma_+] + b^\prime\, H[\Gamma_-], 
\label{eq:bilayer2}
\end{align}
where
\begin{align}
\label{eq:gamma_pm}
\Gamma_\pm &= \frac{1}{2} \left(\lambda_1 + \lambda_2 \pm 
\sqrt{(\lambda_1-\lambda_2)^2 + 4t_\perp^2}\right),\\ 
\lambda_r &=\omega^+ - \epsilon_c - \Sigma_{cr}(\omega),\;\;r=1,2\nnu,
\end{align}
\begin{align}
\label{eq:coeff}
a &= \frac{\Gamma_+ - \lambda_2}{\Gamma_+ - \Gamma_-}
\;\;{\rm and}\;\;b = \frac{\lambda_2 - \Gamma_-}{\Gamma_+ - \Gamma_-}, \\
a^\prime &= -\frac{\lambda_1 - \Gamma_+}{\Gamma_+ - \Gamma_-}
\;\;{\rm and}\;\;b^\prime = \frac{\lambda_1 - \Gamma_-}{\Gamma_+ - \Gamma_-},
\end{align}
and $H[z]$ represents the Hilbert transform with respect to the
non-interacting layer density of states. These equations may now be analyzed 
for a few specific cases. 

\subsubsection{A clean Kondo insulator-metal bilayer system}
As a first application of Equations~\eqref{eq:bilayer1} 
and \eqref{eq:bilayer2}, 
we consider a 
Kondo-insulator-metal interface.
The paramagnetic Kondo insulator is just a renormalized band-insulator, hence
in the particle-hole symmetric limit, $\epsilon_c=0$
 and $\Sigma_c(\omega)=V^2/(\omega^+ - \ep_f
-\Sigma_f)$. In the low frequency limit, the Fermi-liquid form of the 
$f$-selfenergy may be used to get $\Sigma_c \stackrel{\omega\rightarrow 0}{\sim}
ZV^2/\omega^+$, where $Z=\left(1-\partial\Sigma_f/\partial\omega\right)^{-1}$
is the $f$-electron quasiparticle weight. 
Thus, $\lambda_1=\omega^+ - ZV^2/\omega^+$ 
for the Kondo insulator and $\lambda_2=\omega^+$ for the non-interacting metal.
Given these, the c-spectral  functions 
$A_{l}(\omega)=-{\rm Im}G^c_{ii}(\omega)/\pi$ 
for the Kondo insulator ($i=1$, $l=KI$) and metal ($i=2$, $l=M$) 
layers in the low frequency limit are given by,
\begin{align}
  \label{eq:KI-M1}
  A_{KI}(\omega)& \stackrel{\omega\rightarrow 0}{\sim} 
  \left(\frac{\omega t_\perp}{ZV^2}\right)^2  
  \rho_0\left(\omega\left[1+\frac{t_\perp^2}{ZV^2}\right]\right)\\ \nonumber
  &+ \left(1 - \left(\frac{\omega t_\perp}{ZV^2}\right)^2\right) 
  \rho_0\left(\omega\left[1-\frac{t_\perp^2}{ZV^2}\right] 
  - \frac{ZV^2}{\omega}\right),
 \end{align}
\begin{align}
  \label{eq:KI-M2} 
 A_{M}(\omega) \stackrel{\omega\rightarrow 0}{\sim} 
  &\left(1 - \left(\frac{\omega t_\perp}{ZV^2}\right)^2\right) 
  \rho_0\left(\omega\left[1+\frac{t_\perp^2}{ZV^2}\right]\right)\\ \nonumber
 &+ \left(\frac{\omega t_\perp}{ZV^2}\right)^2 
  \rho_0\left(\omega\left[1-\frac{t_\perp^2}{ZV^2}\right] 
  - \frac{ZV^2}{\omega}\right).
\end{align}
From the above expressions, we see that the presence of interlayer coupling
$(t_\perp >0 )$ leads to a linear mixing of the  Kondo-insulator and 
metallic layer spectra in all layers. In $A_{KI}(\omega)$, 
the first term contributes a quadratically vanishing spectral weight into
the hybridization gap, while the second-term leads to the usual gapped spectrum 
of the Kondo insulator. In $A_{M}(\omega)$, 
the first term implies that the spectrum
of the non-interacting metal is strongly renormalized by the proximity to 
the Kondo insulator, and a Kondo resonance like feature must emerge 
in the vicinity of the Fermi level. The second term is gapped at the 
Fermi level, but should lead to a step like feature at $\omega \sim ZV^2$. 
Thus, the metallic states tunnel into the gap of the Kondo insulator 
giving rise to a quadratically vanishing gap, while the
strongly correlated Kondo insulator tunnel into 
the non-interacting metal leading to strong renormalization of the 
non-interacting spectrum.  In the Kondo insulator, 
the $f$-electron spectrum is related to the $c$-electron spectrum through 
$A^f_{KI}(\omega)=(Z^2V^2/\omega^2) A_{KI}(\omega)$ in the limit 
$\omega\rightarrow 0$. 
Due to the tunneling of the metallic states into the Kondo insulator, 
the $f$-spectrum thus becomes gapless. These proximity effects 
are quite general and only assume a linear expansion in frequency of the real 
part of the $f$-selfenergy. Similar results have been observed numerically 
in a recent study on a single Kondo 
insulator embedded in a three-dimensional non-interacting  
metallic host \cite{Peters1} and in a 
theoretical analysis of the surface density of states of heavy fermion 
materials \cite{Peters3}. This physical scenario of "Kondo Proximity Effect" 
was used to qualitatively explain the experimental surface spectra of 
CeCoIn$_5$ reported by Aynajian \etal \cite{Yazdani2012}. 

\subsubsection{A clean Kondo insulator-band insulator bilayer system}
\label{sec:bilayer}
Next, we consider a Kondo-insulator-band-insulator interface. For the band insulator, 
$\lambda_2=\omega^+ - V^2/\omega^+$. Thus, the spectral functions are again 
given by  a linear mixing of the Kondo and the band insulator spectra. A low frequency 
analysis similar to the one carried out above for the Kondo insulator-metal system leads to 
the following. The gap in the Kondo insulator is well known to be substantially 
reduced compared to the band insulator due to the exponentially small quasiparticle weight 
factor $Z$ arising in the strong coupling limit.  Thus, from our analysis, we 
see that the Kondo insulator spectrum close to the Fermi level remains almost unchanged. 
However, in the frequency region where the band insulator had a gap,
but the Kondo insulator had states, a quadratically vanishing spectral weight
tunnels in from the Kondo insulator layer into the band insulator. 

\subsubsection{A dirty Kondo insulator-metal bilayer system}
In this sub-section, we consider, in detail, a substitutionally 
disordered-Kondo-insulator
interfaced with a non-interacting metal. The random substitution
of $f$-sites in the Kondo insulator-layer leads to Kondo hole disorder, 
which is known to lead to an impurity band at the Fermi level
\cite{Schlottmann, *Risebourough, *Tetsuya, Schlottmann2}. 
Specifically, $\lambda_1$
becomes complex because of the static contribution to the selfenergy from 
scattering by impurities \cite{CPA_NFL}. 
Thus $\lambda_1 = \omega^+/Z + i\Gamma_0$, where
$Z$ is the local $f$-electron quasiparticle weight and 
$\Gamma_0=-{\rm Im}\Sigma^{\CPA}_c(0)$
is the scattering rate at the Fermi level.
In a previous paper \cite{CPA_NFL}, we showed that
$\Sigma^{\CPA}_c$ may be related 
to $\Sigma_c$ through a cubic equation if $\rho_0(\epsilon)$ has a 
semi-elliptic form that corresponds to an infinite-dimensional Bethe lattice.
This equation, if used for a single isolated Kondo insulator, yields that
\begin{equation}
\Sigma^{\CPA}_{c}(0)= -i t_{||}\frac{1-p}{2\sqrt{p}}\,,
\label{eq:sigcpa0}
\end{equation}
Thus, at the Fermi level, the conduction electrons acquire a 
selfenergy that is non-analytic in the concentration of 
Kondo holes, $p$.
It was also shown in Ref.~\onlinecite{Schlottmann2} that the $f$ electron 
selfenergy acquires a $1/\sqrt{p}$ dependence when $p\to0$.
However, in the presence of
interlayer coupling, $\Gamma_0$ becomes a complicated function of 
$t_\perp$ and $t_{||}$ and we have not been able to find a simple closed form 
expression like Equation~(\ref{eq:sigcpa0}) for the bilayer case. Nevertheless,
numerical calculations lead us to the conclusion that $\Gamma_0$ 
does retain the same form as the isolated layer case. 

Our natural focus is on the strong coupling regime of the
Kondo insulator, where the quasiparticle weight becomes exponentially
small, i.e $Z\rightarrow 0$. In such a situation,
and with $\lambda_1 = \omega^+/Z + i\Gamma_0$ and $\lambda_2 = \omega^+$,
the values of $\Gamma_\pm$ in Equation~(\ref{eq:gamma_pm}) may be 
simplified to
\begin{equation}
\Gamma_\pm \stackrel{Z\rightarrow 0}{\longrightarrow} \frac{1}{2}\left(
\lambda_1 \pm \sqrt{\lambda_1^2 + 4t_\perp^2}\right)\,.
\end{equation}
The coefficients of the Hilbert transforms in 
equation~\eqref{eq:bilayer1} and \eqref{eq:bilayer2} are given by
\begin{align}
& a\rightarrow  \frac{\Gamma_+}{\sqrt{\lambda_1^2+4t_\perp^2}}\;\;\;
b\rightarrow \frac{-\Gamma_-}{\sqrt{\lambda_1^2+4t_\perp^2}}\nnu \\
&a^\prime=b\;\;{\rm and}\;\;b^\prime=a\,.
\end{align}

These equations are quite easy to analyze at the Fermi level ($\omega=0$). 
In fact, the results from this analysis do not assume strong coupling, 
because if we substitute
$\omega=0$ in the above expressions, the quasiparticle weight does not appear
in them. Hence the following results are valid for any 
coupling strength. 
At the Fermi level, the eigenvalues are given by
\begin{equation}
\Gamma_\pm = \frac{i}{2}\left(
\Gamma_0 \pm \sqrt{\Gamma_0^2 - 4t_\perp^2}\right)\,.
\end{equation}
$\Gamma_0$ is a monotonically decreasing function of the Kondo hole 
concentration (see equation~\eqref{eq:sigcpa0}) and hence in the dilute 
limit ($p\rightarrow 0$), the eigenvalues will be purely imaginary; with 
increasing $p$, $\Gamma_\pm$  become degenerate at $\Gamma_0=2t_\perp$, 
which translates to a specific concentration, $p_d$. In general, $p_d$ 
is a complicated function of $\Gamma_0$. However, we have numerically 
verified that when the ratio $t_\perp/t_{||}\ll 1$, Eq.~\eqref{eq:sigcpa0} 
can be used as an estimate of $p_d$ for multi-layer systems. For such a 
regime $p_d=(\sqrt{1+(2t_\perp/t_{||})^2} - 2t_\perp/t_{||})^2$.
For $p>p_d$, the
two eigenvalues have the same imaginary part but differ in the real part. 
It turns out that the density of states at the Fermi level can be easily 
obtained in two limits: the dilute limit ($p\rightarrow 0$) and the 
degenerate limit ($\Gamma_0=2t_\perp$ or $p=p_d$). We proceed to obtain these.

In the dilute limit, $\Gamma_0 \gg t_{||},t_\perp$, 
so $\Gamma_+\rightarrow i(\Gamma_0 - t_\perp^2/\Gamma_0)$ 
and $\Gamma_-\rightarrow  it_\perp^2/\Gamma_0$.
For purely imaginary arguments ($z=i\eta$), the Hilbert transform with 
respect to a
semi-elliptic density of states may be written in closed form as
\begin{equation}
{\cal{H}}(i\eta)=\int\,\frac{\rho_0(\ep)}{i\eta-\ep} = -\frac{2\,i}{t_{||}}
\left[\sqrt{1+{\bar{\eta}}^2} - {\bar{\eta}}\right],
\label{eq:ht}
\end{equation}
where ${\bar{\eta}}=\eta/t_{||}$. So the Hilbert transform is also purely 
imaginary, which can be expected in the particle-hole symmetric limit. Using 
the above Hilbert transform result and the simplification of the eigenvalues 
in the dilute limit, the density of states of the Kondo insulating layer 
and the non-interacting metallic layer are given by
(to ${\cal{O}}(1/\Gamma_0^2)$)
\begin{align}
A_{KI}(\omega=0) &\stackrel{p\rightarrow 0}{\longrightarrow} 
\frac{1}{\pi\Gamma_0}\left[ 1
-\frac{2t_\perp^2}{\Gamma_0t_{||}}\right] \label{eq:dki1}\\
A_{M}(\omega=0) &\stackrel{p\rightarrow 0}{\longrightarrow} 
\frac{2}{\pi t_{||}}\left[ 1 
-\frac{t_\perp^2}{\Gamma_0 t_{||}} + 
\frac{t_\perp^2}{\Gamma_0^2}\right] \label{eq:dki2}\,.
\end{align}
Note that in all the equations above and below, $A_{KI}(\omega)$ 
denotes the $c$-electron spectral function of the Kondo insulator. 
The above equations reveal the proximity effect of disorder on the two layers.
In the absence of inter-layer tunneling, the impurity band in the Kondo 
insulator grows as $\sqrt{p}$ with increasing Kondo hole substitution 
(since $\Gamma_0\sim1/\sqrt{p}$ in the dilute limit); and the metallic layer 
has a fixed density of states ($=2/(\pi t_{||})$) at $\omega=0$. 
When $t_\perp$ is turned on, the impurities introduced in the Kondo insulating 
layer affect the density of states of the non-interacting clean 
metallic layer, and the relative change in the density of states with respect 
to the isolated layers ($t_\perp=0$) case
is $\sim \sqrt{p}$ for both layers.

In the degenerate limit, $\Gamma_0=2t_\perp$, the eigenvalues are degenerate,
and Eq.~\eqref{eq:bilayer1} and \eqref{eq:bilayer2} cannot be used. 
We revert back to the basic equation~\eqref{eq:finalgc} and after a bit of 
algebra, obtain the density of states of the two layers as 
\begin{align}
A_{KI}(\omega=0) &\stackrel{p=p_d}{\longrightarrow} \frac{2}{\pi t_{||}}
\frac{\left(t_r - \sqrt{1+t_r^2}\right)^2}{\sqrt{1+t_r^2}}, \\
A_{M}(\omega=0) &\stackrel{p=p_d}{\longrightarrow} 
\frac{2}{\pi t_{||}}\frac{1}{\sqrt{1+t_r^2}},
\end{align}
where $t_r=t_\perp/t_{||}$ is the ratio of the inter-layer to 
intra-layer hopping.
We have seen that the metallic layer acquires a Kondo resonance at the 
Fermi level due to the proximity to the Kondo insulator.
Since in the concentrated limit ($p\rightarrow 1$), the entire system 
simply becomes a non-interacting metallic bilayer, the density of states 
must approach the value of $\rho_0(t_\perp)$. It can also be shown that, 
the $A_{M}(\omega=0)>\rho_0(t_\perp)$.
This implies that both the metallic layer and the Kondo insulating layer 
experience a monotonic change in the density of states with increasing Kondo 
hole concentration; the Kondo insulating layer steadily transforming into a 
single-impurity system. 

\subsubsection{Symmetric tri-layer systems}
Another class of systems, for which closed form expressions may be 
readily obtained is a set of three layers in which the outer two layers are 
identical in all respects while the middle layer is different. The
c-Green's function for such a system (denoted by $1-2-3$) is given by
\begin{align}
  G_{11}(\omega)&= G_{33}(\omega)=\frac{1}{2} H[\lambda_1] 
  - t_\perp^2\left( a\,H[\Gamma_+] +
  b\,H[\Gamma_-]\right),\label{eq:tril1} \\
  G_{22}(\omega)& = a^\prime\, H[\Gamma_+] + b^\prime\, H[\Gamma_-], \\
  {\rm where} & \label{eq:tril2} \\
  \Gamma_\pm &= \frac{1}{2} \left(\lambda_1 + \lambda_2 \pm 
  \sqrt{(\lambda_1-\lambda_2)^2 + 8t_\perp^2}\right) \nnu, \\
  \lambda_r &=\omega^+ - \epsilon_c - \Sigma_{cr}(\omega) \nnu, \\
  a&=\frac{1}{(\lambda_1- \Gamma_+)(\Gamma_+ - \Gamma_-)} 
  \;\;{\rm and}\nonumber \\
  b&=\frac{-1}{(\lambda_1- \Gamma_-)
  (\Gamma_+ - \Gamma_-)} \nnu, \\
  a^\prime &= -\frac{\lambda_1 - \Gamma_+}{\Gamma_+ - \Gamma_-}
  \;\;{\rm and}\;\;b^\prime = \frac{\lambda_1 - \Gamma_-}
  {\Gamma_+ - \Gamma_-}.
\end{align}
For the trilayer system, if the layers are uncoupled ($t_\perp=0$) then
the Green's functions should be simply given by $G_{ii}(\omega)=H[\lambda_i]$.
In order to approach the uncoupled layers limit, in the 
equations above, the limit of $t_\perp \rightarrow 0$ 
must be taken with care. We have investigated the symmetric tri-layer system 
in several situations, that are similar to the ones discussed for the bilayer 
system. One of these is
a Kondo insulator, which could be clean or substitutionally Kondo hole 
disordered, sandwiched between two metallic or band-insulating layers. 
The conclusions reached in such cases were
found to be qualitatively very similar to those in the bilayer systems, hence
we now proceed to larger systems and investigate the penetration of disorder
into several clean, non-interacting or interacting layers.

In the next sub-section, we discuss full numerical solutions using the formalism described in 
Section~\ref{sec:model}.

\subsection{Numerical results}
\begin{figure}[h!]
  \centerline{\includegraphics[clip=,scale=0.4]
                        {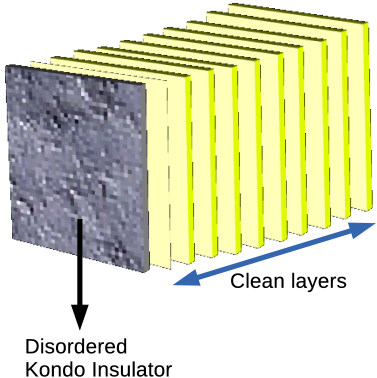}}
\caption{Schematic of the geometry considered in this work. While the
boundary layer is chosen to be a dirty Kondo insulator, the proximal
layers are clean. Three distinct possibilities for these
clean layers have been investigated, namely band-insulators,
metals or Kondo insulators.} 
\label{fig:schem}
\end{figure}

It is evident that the spectral changes in the $c$- and $f$-electron 
Green's functions are a combined effect of the three physical 
parameters in the problem, 
namely: (i) interlayer-hopping, $t_\perp$, (ii) interaction, $U$, 
and (iii) Kondo hole concentration ($p$). In order to disentangle the 
sole effect of disorder from the results, we first discuss briefly 
the non-interacting and clean layered systems. 
Then, we add interactions and note the combined effect of
inter layer coupling and interactions. Finally, we add disorder and by 
comparing the obtained spectra to those of the non-disordered case, 
we isolate the proximity effects of disorder, which represents the main 
objective of this paper.
We study three different cases, with a single substitutionally 
disordered Kondo insulator layer next to (a) several band-insulator, (b) several uncorrelated metal
and (c) several clean Kondo insulator layers. These three cases will be referred to
as disordered Kondo insulator-band insulator, disordered Kondo 
insulator-metal
and disordered Kondo insulator-Kondo insulator, respectively. A 
schematic of the geometries is 
shown in Fig.~\ref{fig:schem}. The number of clean layers has been varied 
from 1 to 11. 
The clean metals are just $U=0$ layers with a simple tight-binding 
Hamiltonian. In the disordered Kondo insulator-Kondo insulator case, 
all the Kondo-insulating layers including the disordered layer have 
the same $U$ and $V$. 
All of the numerical results have been obtained by using a 
semi-elliptic bare density of states for each layer, namely, 
$\rho_0(\epsilon)=\sqrt{1-\epsilon^2/t_{||}^2}/(2\pi t_{||})$, 
with $t_{||}=1$ as the unit of energy.
We begin this section by considering the effects of 
$t_\perp$ only, in a clean, non-interacting, 
$f$- conduction-electron system, interfaced with several non-interacting  metals or several other 
non-interacting  $f$- conduction-electron systems. 
\subsubsection{Effects of inter-layer hopping: $t_\perp\ne0$, $U=0$, $p=0$}
The exact solution of a system of $N$-identical layers discussed in the 
previous section shows that Friedel oscillations are induced in the density 
of states when the bulk Fermi surface is perturbed. For an infinite layer 
system, the inter-layer hopping would simply create an energy band in the 
direction perpendicular to the planes. The introduction of a terminal surface 
into such a bulk system gives rise to surface Friedel oscillations. For finite 
layer-systems, the surface Friedel oscillations would be very pronounced and 
would naturally depend on the nature of the terminal/boundary layer. 
For non-identical layers, exact solutions are not available, however, the 
numerical solution may be obtained almost trivially.
\begin{figure}[h!]
  \centerline{\includegraphics[clip=,scale=0.6]
              {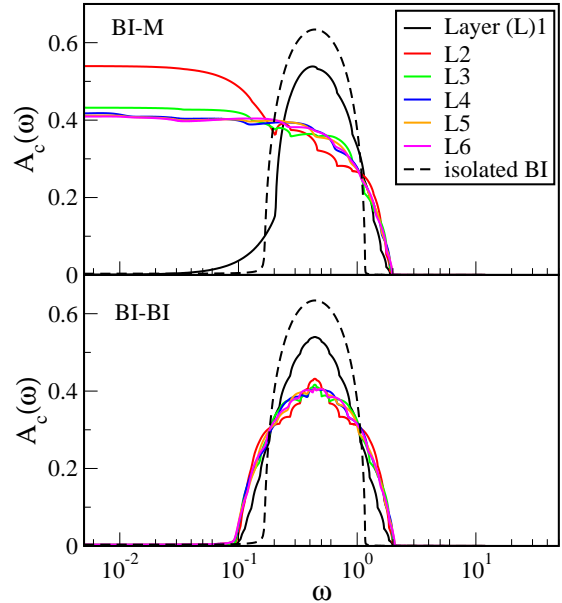}}
  \caption{Spectral density of the $c$ electrons in the presence of non-zero 
           interlayer coupling for 
           a single band insulator layer (L1) at one end of a 11-layer (L2-L12)
           metallic system (top panel) and 
           a single band insulator at one 
           end of a 11-layer band insulating system (bottom panel). 
           The model parameters are 
           $t_\perp=0.5,\, U=0,\, 
           V=0.44$. Note that only $\omega >0$ is displayed on a 
           logarithmic axis. For comparison, the spectrum of an isolated 
           band insulator 
           layer is shown as a dashed line.}
  \label{t_perp_effect}
\end{figure}  
In Fig.~\ref{t_perp_effect}, the layer-resolved $c$-spectra
for a single band insulator coupled to $11$ metallic layers is shown. Since the entire system is particle-hole symmetric, we 
choose to display only the $\omega >0$ spectrum on a logarithmic axis. 
The spectrum for an isolated band insulator layer is shown as a dashed line in both
panels. From the top panel of figure~\ref{t_perp_effect}, we observe that 
the presence of $t_\perp$ between a metallic layer next to a band insulator leads to the 
tunneling of metallic states into the otherwise large hybridization gap of 
the band insulator. We had predicted earlier, through 
equations~\eqref{eq:KI-M1} 
and ~\eqref{eq:KI-M2}, that the presence of $t_\perp$ leads to a quadratic 
rounding off of the band insulator band edge, which is evident when compared 
with the 
hard band edge of the isolated layer's hybridization gap (dashed line). 
The multiple Friedel oscillations due to a non-zero $t_\perp$ are also 
visible in Fig.~\ref{t_perp_effect}. Although equations~\eqref{eq:KI-M1} 
and ~\eqref{eq:KI-M2}  were exact for a bilayer system, the layer resolved 
spectra follow the same qualitative changes in this multi-layered system. 
These oscillations naturally attenuate sharply in the layers that lie far 
from the boundaries. However, for all the finite-layer
systems that we have studied, these oscillations are present. Nevertheless, 
as Freericks \etal discussed in Ref.~\onlinecite{IDMFT2004}, and also found by us, 
the amplitude of these oscillations 
in the surface layers become frozen-in once the system becomes 
large ($\gtrsim$ 10 layers in practice). 

Next, we explore the combined effect of inter-layer coupling and interactions.
\subsubsection{Combined effect of inter-layer hopping and interactions:
$t_\perp\ne0$, $U\ne0$, $p=0$}
\begin{figure}[ht!]
  \centerline{\includegraphics[clip=,scale=0.6]
              {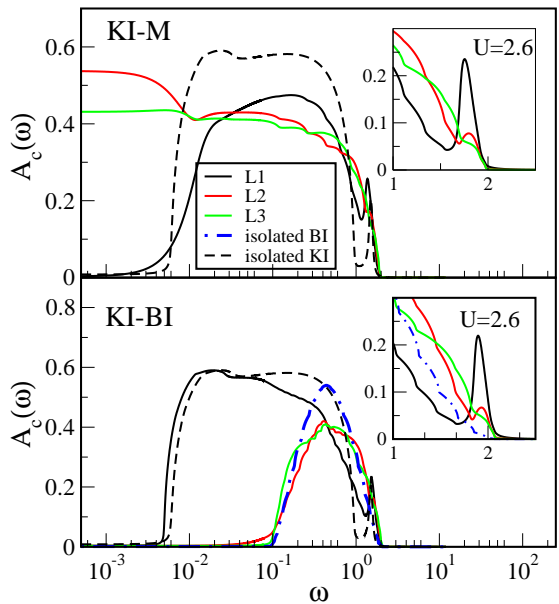}}
  \caption{Layer resolved spectral function for the $c$ electrons 
           due to non-zero interlayer hopping and interactions. 
           A single Kondo insulator layer (L1) is coupled 
           to a 11 layer metallic system (top panel) and a 
           11 layer band insulating system (bottom panel). The spectra for 
           the nearest (L2) and second nearest (L3) metallic (top panel) and 
           the band insulator (bottom panel) is shown.
           For both panels inter-layer hopping 
           $t_\perp=0.5$, and $c-f$ hybridization is set to $V=0.44$ 
           for both the Kondo insulators and the band insulators, 
           Coulomb repulsion in the Kondo insulating layer $U=1.7$ 
           in the main panels and $U=2.6$ in both insets. 
           For comparison, the $c$-DOS of an isolated Kondo insulator and 
           an isolated band insulator is also shown as (black) dashed 
           (top and bottom panel) and (blue) dashed-dotted lines 
          (bottom panel), respectively. The insets show the Hubbard bands 
          that emerge as a result of their proximity to the coupled Kondo 
          insulator. This effect is visibly detectable only until the second 
          nearest neighbor layer.}
  \label{t_perp_U_effect}
\end{figure}
 
In the context of bulk systems, the presence of interactions on the $f$-orbitals
 in an otherwise simple band insulator leads to a strongly renormalized 
hybridization gap and hence a Kondo insulator. The proximity of a strongly 
interacting Kondo insulator to a non-interacting layer, either a band 
insulator or a metal, is an interesting problem.  Our analysis 
in previous section showed that the proximity of a 
band insulator to a Kondo insulator layer results in a rounding off of the 
hybridization band edge of the band insulator, while the Kondo insulator 
layer remains almost unaffected.
If the band insulator is replaced by a metal, then we saw that the 
metallic states tunnel into the Kondo insulator, while strong correlations 
in the Kondo insulator tunnel into the non-interacting
metal and give rise to a Kondo resonance like feature. This is consistent with 
the numerical renormalization group results of 
Peters \etal on Kondo superlattices \cite{Peters1}.
Our results, although derived for bilayer 
(Kondo insulator-band insulator and Kondo insulator-metal) systems, remain
robust even when the number of band insulator or metal layers is increased. 
For example, in Fig.~\ref{t_perp_U_effect} 
we show the layer-resolved $c$-spectra for a single Kondo insulator 
and 11 metallic layer system (top panel) and a single 
Kondo insulator and 11 band insulator layer  system  
(bottom panel). 
In this bottom panel, due to the presence of finite density of states in the 
proximal Kondo insulator, we see that the $c$-spectra of the  second neighboring 
band insulator layer acquires a quadratic rounding off around the band edge. 
The farther band insulator layers are however inert to these effects.

The top panel of Fig.~\ref{t_perp_U_effect} shows that, for a metallic 
layer adjacent to a Kondo insulator, the proximity
effect results in a Kondo-like resonance 
at the Fermi level of the otherwise non-interacting metal. 
This is the Kondo proximity effect
discussed in Ref.~\onlinecite{Peters1}. 
Proximity to a strongly interacting Kondo insulator induces changes not only at low energies such
as the Kondo scale, but also
at high energies such as Hubbard bands. The adjacent non-interacting band-insulating or metallic layers also acquire 
minute Hubbard bands indicating the tunneling of electron correlations into the 
non-interacting layers. Similar effects 
occur even in bulk heavy fermions, modeled by the periodic Anderson model, where although the $U$ is present 
only for $f$-electrons, the mixing between $c$- and $f$-electrons through the hybridization produces Hubbard 
bands in the spectrum of the conduction band electrons. 
Whether the single Kondo insulator is coupled to band-insulating or metallic layers,
these high energy proximity effects are the same, which is evident from the insets of Fig.~\ref{t_perp_U_effect}.  
Now we proceed to explore the spectral modifications induced in the layer 
resolved spectra when we introduce Kondo holes in the boundary Kondo insulator layer.
\subsubsection{Effects of Kondo hole disorder in the boundary layer}
\begin{figure}[h!]
  \centerline{\includegraphics[clip=,scale=0.5]
                        {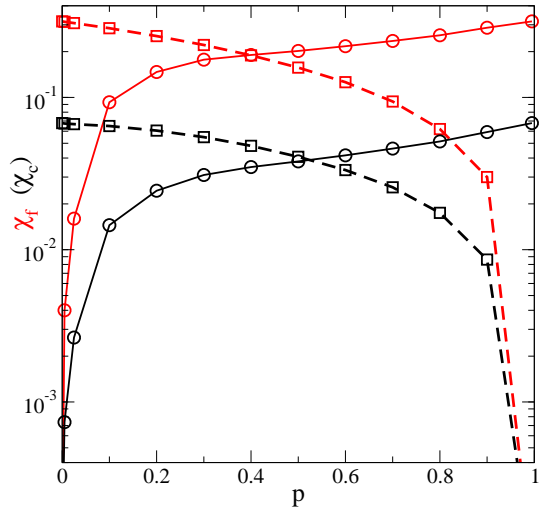}}
  \caption{The spectral weight difference between the disordered
  and the clean $c-$ and $f-$spectral functions for a single PAM layer as 
  a function of Kondo hole concentration $p$. Red lines and symbols 
  correspond to the $f$ spectra function, while
  black lines and symbols correspond to the  $c-$spectral functions. 
  The solid lines with 
  circles represent the spectral weight difference with respect to 
  the $p=0$ (clean) case, 
  the dashed lines with squares correspond to the comparison with the 
  $p\rightarrow 1$ limit. The parameters used are $U=1.7$, 
           $V=0.44$, and $t_\perp=0.0$.} 
  \label{fig:sdKI}
\end{figure}
The effects of Kondo-hole disorder in the bulk PAM have been extensively 
investigated \cite{Schlottmann, *Risebourough, *Tetsuya, 
Schlottmann2, CPA_NRG, CPA_NFL}. 
It is well known that there is a crossover from coherent lattice behavior to incoherent 
single-impurity behavior as the disorder $p$ changes from zero to one \cite{CPA_NRG, CPA_NFL}. 
Such a crossover is reflected in all physical
quantities, including resistivity, thermopower and density of states.
For layered systems, the range and extent of disorder effects may be quantified through
a  measurement of integrated spectral weight difference in a given
frequency interval $|\omega| \leq \lambda$. 
We define
$\chi_\nu(p,p_0;\alpha;\lambda)$, as the spectral weight difference of the $\alpha^{\rm th}$ layer, computed through
\begin{align}
 &\chi_\nu(p;p_0;\alpha;\lambda)\nonumber\\
 &=\int^\lambda_{-\lambda} \,d\omega\,\left|A\left(p;\{U_\beta\};\alpha;\omega\right)
-A(p_0;\{U_\beta\};\alpha;\omega)\right|
\label{eq:chi}
\end{align}
for a set of fixed interaction
strengths $\{U_\alpha\}$. Here $\nu=c/f$ and $A(p;\{U_\beta\};\alpha;\omega)=-{\rm Im}G_{\alpha}(p;\{U_\beta\};\omega)
/\pi$ is the spectral function of the $\alpha^{\rm th}$ layer's 
when the Kondo hole concentration is $p$. Physically, this quantity represents the extent to which the layer density
of states changes when disorder goes from $p_0$ to $p$ or vice-versa when
all other parameters are fixed.
We have employed $n_c=1$ and $n_f=1$  for an isolated Kondo insulator layer.
With random dilution of $f$-sites, which introduces Kondo hole disorder, an impurity
band is expected to form at the Fermi level. The spectral weight difference of an isolated disordered Kondo insulator layer
with respect to the clean ($p_0=0$) and single-impurity ($p_0\rightarrow 1^-$)
limits as a function of
Kondo hole concentration $p$  is shown in Fig.~\ref{fig:sdKI} (with the cutoff
$\lambda\rightarrow \infty$).
The spectral weight difference between the disordered and the clean case integrated over all frequencies 
(shown in Fig.~\ref{fig:sdKI} as solid lines with circles) is seen to rise rapidly as a function of increasing $p$ and 
roughly saturate at the 
concentration of $p\sim 0.5$.
The spectral weight difference at $p\sim 0.5$  is large when compared with either  
 clean case ($p=0$)  or the $p\rightarrow 1^-$
single-impurity case (shown in Fig.~\ref{fig:sdKI} as dashed lines with squares). 
Hence we choose  
a fixed disorder $p=0.5$ on the boundary layer in all of our subsequent discussion, unless otherwise mentioned.
The rest of the layers are chosen to be clean. 

With the experimental realizations of CeIn$_3$/LaIn$_3$ 
superlattices~\cite{dimensional_confinement2010} and
thin films of non-magnetic Yb substituted into 
CeCoIn$_5$~\cite{CeCoIn_thin_films}, 
the possibility of having an interface between a  Kondo hole disordered layer
and a clean layer cannot be ignored. 
We focus on the following questions: How does the random dilution of
$f$-electrons in a single Kondo insulator boundary layer 
(see schematic Fig.~\ref{fig:schem})
 affect the dynamics of the electrons in the adjoining clean layers? How far
 do the disorder effects penetrate? And does this range get enhanced or suppressed
 by the presence of interactions?

In order to 
answer these questions, we therefore investigate the interface of a single 
disordered Kondo insulator and several clean layers that could be non-interacting metals or band-insulators
or strongly interacting Kondo insulators. 
We start our discussion with a system of 12 layers where the 
boundary layer is a disordered Kondo insulator and the rest
of the layers are non-interacting metals. We refer to this as a disordered Kondo insulator-metal system. 
\begin{figure*}[htb]
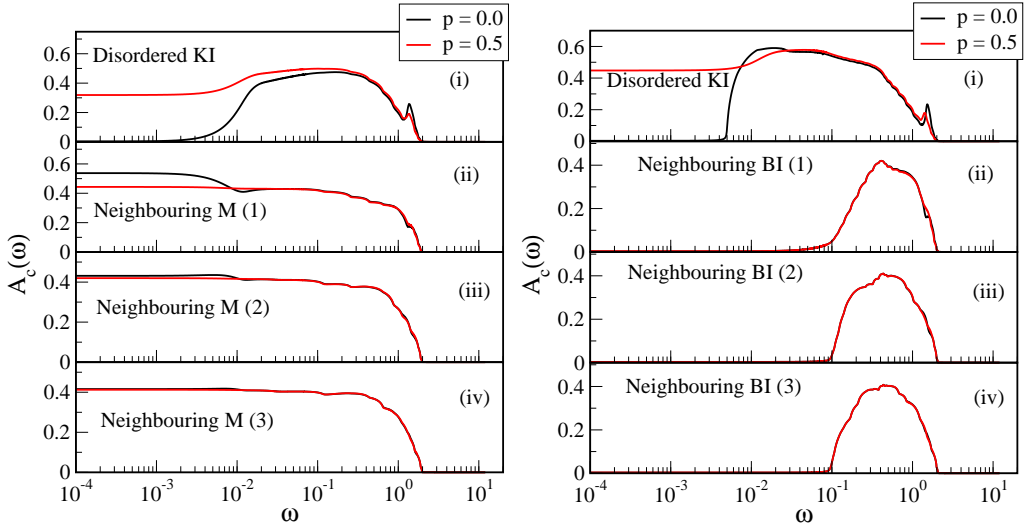

  \centerline{\includegraphics[clip=,scale=0.5]
             {1DKI-11M_c.eps}
             {\includegraphics[clip=,scale=0.5]
              {1DKI-11BI_c.eps}}}
  \caption{The $c$-electron spectra of the respective layers as 
  denoted in the panels for a single disordered Kondo insulator interfaced 
  with a 11-layer metallic (M) (left panel) or band insulating (BI)
  (right panel) system for a clean ($p=0$) (solid black) and a disordered
  ($p=0.5$) Kondo insulating (KI) layer (solid red). For the disordered Kondo 
  insulator this represents the CPA average of $-\Im G_{\mathrm{CPA}}/\pi$.
  The model parameters are $t_\perp=0.5,\,V=0.44,\, U=1.7$.}
\label{dKI-M}
\end{figure*}
 The proximity of a clean Kondo insulator to a metal leads to (i) 
 a quadratic band bending at the Kondo hybridization band edge of the Kondo insulator layer
 and (ii) appearance of a Kondo resonance and a strongly renormalized low 
energy spectra induced by the "Kondo  proximity effect". 
This has already been discussed earlier for a bilayer 
 Kondo insulator-metal interface. In the presence of disorder, it is well known that Kondo holes introduce 
 an impurity band in the center of the Kondo hybridization gap 
 \cite{Schlottmann, *Risebourough, *Tetsuya}. This impurity band then fills up the gap 
with increasing disorder concentration, leading to a continuum of metallic 
states spanning the entire gap. This observation holds here too as seen in 
the panel (i) of Fig.~\ref{dKI-M}. This phenomenon due to Kondo holes 
manifests itself in both the CPA averaged $c$-electron spectra and the 
local (impurity) $f$-electron spectra of the disordered 
Kondo insulator (not shown here). Additionally, the Hubbard bands get 
depleted due to the presence of Kondo holes. 
The important aspect however, is that these disorder effects do not just 
remain confined to the disordered layer but also penetrate into the 
neighboring clean layers depicted in the lower panels of figure~\ref{dKI-M}. 
Disorder in fact destroys the proximity effects of interactions. The Kondo
resonance in the adjoining metallic layer at zero disorder appreciably 
reduces in intensity when the Kondo hole concentration, $p$ is $0.5$ and 
tends to disappear as $p$ increases. 
This is more clearly visible from figure~\ref{dKI-M-LE}. 
The low energy spectra of the farther neighbors also acquire 
visible low energy spectral changes that evolve with change in disorder 
concentration, tending to 
crossover to the noninteracting limit of a clean metal/metal interface, as 
seen from figure~\ref{dKI-M-LE}. 
\begin{figure}[htb]
  \centerline{\includegraphics[clip=,scale=0.5]
             {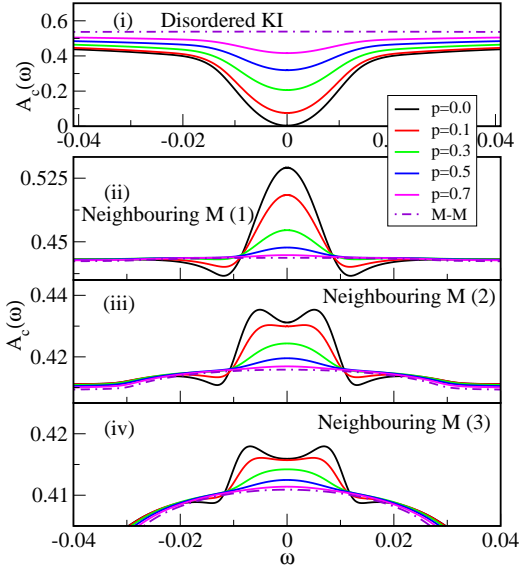}}
\caption{Spectral evolution at low energies for the disordered Kondo insulator 
(panel(i)) and the proximal clean metals (panel((ii)-(iv))). 
A system of a single disordered Kondo insulator and 11 clean metallic 
layers is used.The respective layer-resolved metallic spectra 
corresponding to a 12 layer metal-metal (M-M) system is also shown 
for comparison. The model parameters are the same as used in Fig.~\ref{dKI-M}}.
\label{dKI-M-LE}
\end{figure}
The tiny Hubbard bands in the non-interacting layers also disappear with 
increasing disorder. This can be expected qualitatively because introducing 
Kondo holes implies that sites with $f$-orbitals in the boundary layer are 
randomly being replaced by non-interacting sites, hence the effects
of interactions should get mitigated not only in the layer but also in the 
adjoining layers. 

The penetration of Kondo holes into proximal band insulating layers is 
however different (than the proximity to metals discussed above). 
The exponentially larger non-interacting hybridization gap of the proximal 
band insulators prohibit electrons from tunneling into the insulating gap. 
Thus, disorder induced spectral changes in proximal band insulators remain 
confined to the 
band edges and the high energy Hubbard bands. To this end, we refer back 
to our discussion on the Kondo insulator/band insulator bilayer interface 
in 
section 3.3.1.4. 
There we realized that, around the band 
edges of the proximal band insulator, where the Kondo insulator has available 
states, a quadratic band bending would occur owing to the interlayer 
coupling. If the clean layers are also interacting Kondo-insulators, 
the proximity effects of disorder span over all frequency scales, 
from the universal, hybridization gap scale to the non-universal, 
Hubbard bands. This shall be discussed now. 

\begin{figure*}[t]
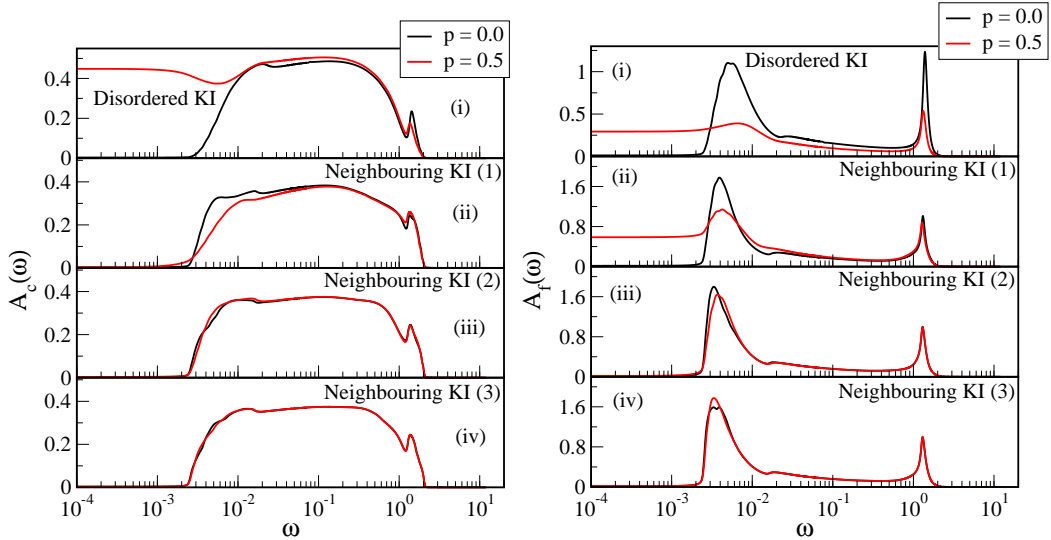

  \centerline{\includegraphics[clip=,scale=0.5]
              {12KI_c.eps}
              \includegraphics[clip=,scale=0.5]
              {12KI_f.eps}}
  \caption{(Left panel): 
            The layer resolved $c$-electron spectra of a single disordered 
            Kondo insulator coupled to 11 clean Kondo insulating (KI) layers.
             The respective layers are denoted in the 
            panels;(right panel): 
            The same for the CPA averaged $f$-electrons of the 
            disordered Kondo insulator and the 
            interacting $f$-electrons of the clean Kondo insulators.
            The parameters used are $t_\perp=0.5,\,V=0.44,\,U=1.7$}
              \label{dKI-KI}
\end{figure*}

The boundary layer metallic states introduced by Kondo hole disorder now tunnel into
the adjoining Kondo insulator, causing the quadratic bending of the gap
in the adjacent layer (see figure~\ref{dKI-KI}). This tunneling effect does propagate into the further layers,
but is attenuated by the $\omega^2$ factor and is hence too small to
be observed.  The mechanism of penetration of the Kondo hole disorder induced metallic states 
from the disordered Kondo insulator to the coupled clean Kondo insulator is thus physically very similar 
to that of a clean metal interfaced with a clean Kondo insulator (Eqs~\eqref{eq:KI-M1}, \eqref{eq:KI-M2}).  

In the disordered Kondo insulator layer, spectral weight transfer occurs across all scales.
It is well known that Kondo hole substitution results in a coherent lattice
to an incoherent single-impurity crossover. This manifests in the transfer of weight from high energy Hubbard bands to the Kondo resonance.
Concomitantly, the hybridization experienced by the interacting sites in
the boundary layer crosses over to a featureless non-interacting lineshape. 
The adjacent layers get strongly affected by these changes
and the spectral weight transfer occurs across all energy scales  in the nearest
neighbor Kondo insulator as well. The Hubbard bands in the second layer too acquire an explicit Kondo hole 
induced depletion. Although there exists no explicit interlayer 
$f$-$c$ hybridization the $f$-electrons in the clean layers get strongly influenced by the 
presence of Kondo holes in its adjacent layer. This is because 
the $f$-e$^-$s of the proximal clean layers see a self consistently 
determined layer dependent host hybridization function, $S(\omega)$, 
as can be seen 
from Eq\eqref{eq:FSE}. We now quantify
the spectral weight transfers focusing on the role of interactions in the
proximity effects of disorder.

\begin{figure}[t]
  \centerline{\includegraphics[clip=,scale=0.5]
              {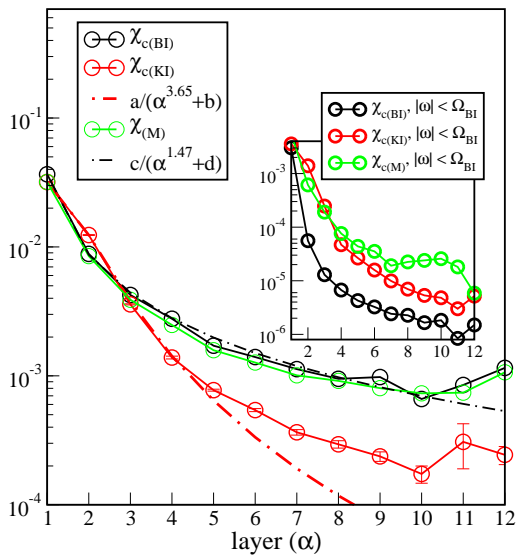}}
  \caption{The integrated spectral weight difference 
           (between the disordered and the clean systems, 
           see equation~\eqref{eq:chi}) as a function of the layer index for 
           a system of a single doped Kondo insulator interfaced with 
          (a) 11 clean band-insulators (black); 
          (b) 11 clean Kondo insulators (red) and 
          (c) 11 metallic layers (green).
          The main panel is the spectral weight difference with a 
          cutoff, $\lambda\rightarrow \infty$.
          In the inset, $\Omega_{\mathrm{BI}}=\frac{1}{2}
          \left(-t_{||}+\sqrt{t_{||}^2+4V^2}\right)$, representing the 
          hybridization gap edge of an isolated band insulator, has been used 
          as the cutoff ($\lambda$) in energy space to get the spectral weight 
          difference at `low' energies $(|\omega|<\Omega_{\mathrm{BI}})$. 
          The dot-dashed lines are algebraic decay fits, with black for (a) 
          and (c), and red for (b). $U=1.7$ in the Kondo insulating layers, 
          $V=0.44$ in both the band insulating and Kondo insulating layers, 
          the inter-layer hopping is set to $t_\perp=0.5$.}
\label{SWD}
\end{figure}
As explained earlier, the spectral weight difference between the spectra, with
and without disorder, integrated in a given frequency interval, represents 
one measure of the proximity effect of disorder. We have chosen
two frequency intervals: (i) a `low' frequency interval defined by
$|\omega|\leq \Omega_{BI}$, where $\Omega_{\mathrm{BI}}=
            \frac{1}{2}\left(-t_{||}+\sqrt{t_{||}^2+4V^2}\right)$, represents 
the hybridization gap edge of an isolated band insulator and (ii) the entire
frequency range. This classification into low frequencies and all frequencies
is done to emphasize that the low frequency spectral changes could be 
tiny ($\sim \omega_L\rho_0(0)$, which is exponentially small in strong coupling)
but affect low temperature properties, especially transport, in a major way.
The spectral changes over {\em all} frequencies will show up in photoemission 
and optical properties measurements, and hence are important from a different
perspective.
Since the Kondo holes are introduced only in the boundary layer, the spectral weight
transfers would be maximal there. Moving away from the disordered Kondo insulator 
layer, the changes in the spectra due to the boundary layer disorder must decrease. Indeed,
this expectation is borne out, as seen in
figure~\ref{SWD}. The spectral weight difference defined in 
equation~\eqref{eq:chi} 
decreases sharply with distance from the disordered Kondo insulator layer for all 
the three cases considered (interfacing the disordered Kondo insulator with 
either 11 metals, or 11 band/Kondo insulators) thus far. 
   These spectral weight changes quantify the proximity effects of 
disorder, in the absence or presence of interactions. The fits of the spectral weight difference (dot-dashed lines in 
figure.~\ref{SWD}) show that
the proximity effects decay algebraically with increasing distance from the boundary
layer. Such a decay profile indicates that changes made in the boundary
layer can penetrate quite deep into the system.

   Apart from the overall decay profile, there are subtle aspects of the spectral weight difference that we highlight now.
The main panel in figure~\ref{SWD} shows the spectral weight difference computed over all frequency scales ($\lambda
\rightarrow \infty$). The spectral changes in the non-interacting as well as 
interacting layers are see to be similar up to the third layer, beyond which
the Kondo insulating layers are far less affected than the non-interacting
metallic or band-insulating layers. However, the low frequency spectral weight difference ($\lambda=
\Omega_{BI}$), shown in the inset, presents an entirely different picture.
The band-insulators are seen to be least affected, since the spectral weight difference drops rapidly
by over a decade even for the layer adjacent to the disordered Kondo insulator boundary layer.
In contrast, the metallic and Kondo-insulating layers experience similar 
levels of spectral changes due to the disorder in the boundary layer. 
Thus, when viewed over all scales, interactions indeed screen the
proximity effects of disorder. However, in the neighborhood of the Fermi level,
the presence of a spectral gap makes the band-insulator immune to the changes
in the boundary layer, and the presence or absence of interactions is irrelevant.

\section{Discussion}
For the results in the previous section, it is clear that Kondo hole substitution in a single boundary layer does indeed
affect neighboring layers to varying degrees on different energy scales.
 Although we have considered
finite systems, a few general remarks may be made for infinite, 
periodic structures.
\subsection{$f$-electron Superlattices:}
A superlattice  structure comprises a periodic array of unit cells,
each of which consist of 
a finite number of $(m)f$-$e^-$ and $(n)c$-$e^-$ layers. For example, in 
Ref.~\onlinecite{dimensional_confinement2010}, $(m,4)$CeIn$_3/$LaIn$_3$ 
superlattices were grown and $m$ was tuned. The authors surmised that 
Kondo hole disorder at the Ce/La interface is inevitable. 
As shown in figure~\ref{SWD}, the extent of 
proximity induced disorder effects is appreciable in the neighboring layers. 
Owing to the periodic nature of a superlattice, it is thus expected 
that this proximity effect would be even more pronounced stemming from 
penetration of disorder effects from the adjacent unit cells of the superlattice.
  
\subsection{A scenario for disorder induced non-Fermi liquid:} 

Thus far, we have considered the particle-hole symmetric limit. However,
an interesting situation might emerge by varying the chemical potential
such that the system does not have particle-hole symmetry.
Dynamical effects of impurity scattering even at the single-site mean 
field level 
have been shown (in Ref.~\onlinecite{CPA_NFL}) to lead to a 
non-Fermi liquid form
of the average selfenergy. Thus the Kondo hole disordered layers 
in such an array of $f$ electron layers would show non-Fermi liquid behavior. The non-Fermi liquid 
nature of the Kondo hole disordered layers would thus introduce a further anisotropy in the 
system that would possibly manifest itself in the transport properties of 
these systems. Moreover, if the Ce/La interfaces in the superlattice
unit cell are separated by a larger number of clean layers, the 
non-Fermi liquid effects would be expected to attenuate. Hence a 
non-Fermi liquid to Fermi liquid crossover can be expected simply by increasing the ratio of the number 
of layers to the number of Ce/La interfaces. 
\subsection{Anderson Localization:}
One of the important consequences of disorder, ignored in this work, is that
of Anderson localization (AL).
The CPA employed here to treat disorder effects
is incapable of incorporating AL. The addition of  coherent back-scattering
and deep trap physics beyond CPA should lead to profound consequences for the
adjacent layers. The typical medium -dynamical cluster approximation 
(TMDCA) developed recently \cite{TMDCA}has been found to be an excellent 
approach for obtaining the correct phase diagram of the non-interacting 
Anderson model. 
It would be interesting to extend the TMDCA for interacting models
and explore the interplay of Anderson localization physics and strong 
interactions
in layered systems.

\section{Summary and Conclusions}

In this work, we have employed the inhomogeneous dynamical mean field theory \cite{IDMFT2004} framework to 
obtain self consistent many body solutions for layered Kondo hole 
substituted $f$-electron systems. The substitutional disorder,  treated within 
the coherent potential approximation, was introduced in a single boundary layer,
and the consequent spectral changes in the neighboring layers
were explored.   Three distinct types of clean adjacent layers were considered:
(a) non-interacting metals (disordered Kondo insulator/metals), (b) non-interacting 
band insulators (disordered Kondo insulator/band insulator), and, 
(c) several clean Kondo insulators.

Combining simple analytical expressions for bilayer and trilayer systems with
full numerical calculations using the local moment approach, we have
(a) explained the strong renormalization of the low energy spectra of 
the proximal layers, (b)  spectral interference among the layers in presence of an 
inter-layer coupling $(t_\perp)$, and (c) a mechanism for penetration of
Kondo hole disorder induced  incoherence into clean layers. We 
highlight the differences between three distinct types of interfaces with a 
disordered Kondo insulator layer. We also demonstrate that, in addition to a robust 
low energy quasiparticle peak \cite{Peters1}, proximity effects of 
interactions also manifest through the appearance of minute Hubbard bands 
in the neighboring non-interacting layers.

A finite concentration of Kondo holes 
leads to the formation of an impurity band at the Fermi level in a Kondo 
insulator
\cite{Schlottmann, *Risebourough, *Tetsuya}, thus creating metallic states 
in the otherwise gapped insulator.
In the layered geometry, these disorder induced metallic states can 
further tunnel into 
the immediately neighboring clean Kondo insulator, 
rendering a quadratic bending in its 
Kondo hybridization gap edge. 
This further induces metallic states in the $f$-e$^-$ 
spectra of the the nearest neighbor layer. These disorder induced 
states do propagate into the further layers but the tunneling is attenuated by the $\omega^2$ factor for each layer
and hence is too small to be observed beyond the nearest neighbor layer. Introduction of disorder 
further destroys the proximity effect of interactions. We showed 
that a manifestation of proximity induced interaction effect, 
on the adjacent non-interacting layers, 
is the appearance of tiny Hubbard bands and a proximal Kondo effect induced low 
frequency spectral renormalization. This proximity effect of interactions, on 
the non-interacting layers, is mitigated when they are coupled to a 
disordered Kondo insulator layer.

The presence of disorder induces spectral weight transfer from the high energy 
Hubbard bands to the low energy Kondo scaling regime not only in the disordered 
Kondo insulator but in the adjacent clean layers as well. We have quantified 
the penetration of disorder effects in these interfaces through the spectral 
weight difference function 
(equation~\eqref{eq:chi}. The spectral weight difference is maximum in the 
disordered Kondo insulator and decays algebraically 
with distance from the disordered Kondo insulator layer.

Our analysis implies that, while for thin films, 
the spectral weight difference of the clean layers
is appreciable only until the third neighboring layer; for superlattice geometries, 
this effect of penetration of disorder would be quite pronounced due 
to the occurrence of multiple disordered interfaces. 
 
We foresee a Fermi liquid to non-Fermi liquid crossover in 
the transport properties across disordered interfaces by tuning the number of 
disordered layers. The full study of such a crossover and inclusion of localization
effects beyond the coherent potential approximation will be the subjects of future projects. 

\acknowledgments
{This work is supported by NSF DMR-1237565, NSF EPSCoR Cooperative Agreement No. EPS-1003897 with 
additional support from the Louisiana Board of Regents, and the DOE Computational Materials and Chemical 
Sciences Network (CMCSN) SC0007091.
S.~S. acknowledges the support of CSIR, India. 
S.~S. also acknowledges the hospitality 
of the department of Physics \& Astronomy at Louisiana State University.}
\bibliography{IDMFT_ref}
\end{document}